\documentclass[prl,twocolumn,showpacs,amsmath,amssymb,aps,floats]{revtex4}

\usepackage{graphicx}
\usepackage{dcolumn}
\usepackage{bm}

\def\Black          {}

\def\nnd{\end{document}}

\def\nnb{\nonumber}

\def\be{\begin{equation}}
\def\ee{\end{equation}}

\newcommand{\bea}{\begin{eqnarray}}
\newcommand{\eea}{\end{eqnarray}}
\newcommand{\bwt}{\begin{widetext}}
\newcommand{\ewt}{\end{widetext}}

\def\wsep{ \nnb \\ &&}

\def\eed{\end{document}}

\def\al{\alpha}

\def\m_z{m_{\textrm {Z}}}
\def\al{\alpha}
\def\be{\beta}

\def\de{\delta}
\def\al{{\alpha}}
\def\my#1{ #1 }

\def\sbbkf#1{\bigg ( #1 \bigg )}

\makeatother

\begin{document}

\preprint{ KEK-TH-1042}
\title{Electroweak symmetry breaking and precision data }
\author{ Sukanta Dutta $^{a,\,b}$}
\author{ Kaoru Hagiwara $^{b,\,c}$} 
\author{Qi-Shu Yan $^b$\,\, }
\affiliation{$^a$ SGTB Khalsa College, University of
  Delhi, Delhi-110007, India.}
\author{  }
\affiliation{$^b$ Theory Group, KEK,  Tsukuba,   305-0801, Japan.}
\affiliation{$^c$ The Graduate University for Advanced Studies, 
Tsukuba, 305-0801, Japan.}

\begin{abstract}
We study the impact of LEP2 constraints on the dimensionless coefficients of the electroweak 
chiral Lagrangian on the precision observables using the improved renormalization 
group equations. We find that the current uncertainty in the triple and quartic gauge 
boson couplings can accommodate electroweak symmetry  breaking models  
with $S(\Lambda=1\, \textrm{TeV}) > 0$.

\end{abstract}

\pacs{11.10.Gh, 11.10.Hi, 12.15.Ji, 12.15.Lk }

\maketitle

\renewcommand{\thefootnote}{\arabic{footnote}}
\setcounter{footnote}{0}
%
%
%
%

There have been recent attempts to constrain the 
electroweak symmetry breaking (EWSB) models  
by analyzing  the electroweak $S$, $T$, $W$ and $Y$  
parameters \cite{Barbieri:2004qk}. However, these studies have not 
considered the uncertainty in the triple gauge couplings (TGC) 
from LEP2 and Tevatron.
In this letter, we study implications of the TGC constraints 
from LEP2 and Tevatron on the parameter space of the 
nonlinearly realized electroweak chiral Lagrangian (EWCL) by
taking into account the logarithmic scale dependence of the chiral
coefficients.

Restricting our study on the bosonic sector of the 
electroweak effective field theory we include all 
the operators upto mass dimension four in EWCL  \cite{Appelquist:1993ka}
which contribute to the two, three  and four point functions.
We confine to consider the set of operators consistent with discrete
symmetries, $P$, $T$, and $C$.
In a similar study, Bagger et. al. \cite{Bagger} 
considered operators contributing to two point functions only. 

In the framework of effective field theory, the dimensionless chiral coefficients
of the EWCL, such as the precision parameter $S$ and $T$, depend on the renormalization
scale as 
\begin{eqnarray}
{\cal O}(m_Z)^{\rm exp}&=&{\cal O}(\Lambda )^{\rm New \,Phys.} + \beta_{\cal O} \ln\left(\Lambda\over m_Z\right).
\end{eqnarray}
We evaluate $\beta_{\cal O}$   by including 
all dimensionless chiral coefficients corresponding to  $O(p^4)$ operators in 
our renormalization group equation (RGE) analysis  
and  take into account the bounds of the
TGC from the LEP2 measurements as our input. 
We have extended our earlier study on computation of one loop RGE  
using background field technique for SU(2) case \cite{dhy1}  to  improve
 upon the existing RGE's \cite{Appelquist:1993ka, old}  for  
EWCL and are presented in reference \cite{full}.

Before presenting the $\beta$ functions of two point chiral coefficients, 
we describe the experimental or theoretical bounds of all 
dimensionless chiral coefficients.

Two point function chiral coefficients are extracted 
from data collected in Z factories.
We perform the analysis with 
the three best measured 
quantities $m_\textrm{W}= 80.425\pm 0.038$ GeV, 
 $\sin^2\theta_W^{\textrm{eff}}=0.23147\pm 0.00017$ 
and the leptonic decay width of $Z$, $\Gamma_l =83.984\pm 0.086$ MeV
for the $S,\, T$ and $U$ fitting. The other inputs 
used are $1/\al_{\textrm{em}}(\m_z)= 128.74$, $\m_z=91.18$ GeV, and
$m_{\textrm{t}}=175$ GeV. 
 
The central values with  $1\sigma$ errors of the $S,T,U$ parameters 
are found as
\begin{equation}
\begin{array}{rl}
S = &(-0.06\pm 0.11)\\
T = &(-0.08\pm 0.14)\\
U = &(+0.17\pm 0.15)
\end{array}
\rho_{co.} =\left(\begin{matrix}
1\,\,\, & & \cr 
.9 & 1\,\,\, & \cr 
-.4&-.6& 1\cr
\end{matrix}
\right)
\label{stfit}
\end{equation}
which roughly agrees with \cite{Bagger}.

The fit is based on one loop calculation and performed using 
the procedure in reference \cite{Hagiwara:1998}. 
In order to make a correspondence with the definition of $S-T$ 
in EWCL, we  subtract the contribution
of Higgs boson from the Standard Model (SM )at a reference value 
$m_H^{ref}$ as  given in \cite{Bagger}. The validity of the 
subtraction method is checked by observing the independence of 
Higgs mass in the fit. 
\par  The relations among the $S-T-U$ parameters with the 
chiral coefficients $\al_1$, $\alpha_0$, and $\al_8$ of EWCL  are found to be
\bea
\al_1 (\mu)&=& - \frac{ S(\mu)  }{16 \pi}, \nnb\\
\al_0(\mu) &=& - \frac{\al_{\textrm{EM}}^{} T(\mu) }{2},  \nnb\\
\al_8 (\mu)&=& - \frac{ U(\mu)  }{16 \pi}\,.
\label{stu}
\eea
which are in complete agreement with those in 
\cite{Appelquist:1993ka} provided we take into account  
the sign difference  for $\beta$ parameter accounted 
for the  calculation performed in Euclidean space  \cite{full}.

Since the $S-T-U$ parameters are defined by Z pole data, 
 Eq. (\ref{stu}) is to be read at $\mu=\m_z$, from which
$\al_1(m_Z)$-$\al_0(m_Z)$-$\al_8(m_Z)$ are determined.

The three point chiral coefficients $\al_2$, $\al_3$ 
and  $\al_9$ are extracted from the LEP2 W pair production
measurements. These three chiral coefficients 
are related to the experimental observable 
$\delta k_\gamma$, $\delta k_{\textrm{Z}}$, 
$\delta g^1_{\textrm{Z}}$ \cite{Hagiwara:1986vm} as
\bea
\delta k_\gamma&=& - (\al_1 + \al_8 + \al_2 + \al_3 + \al_9) g^2 \,,\label{tri0} \\
\delta k_Z &=& - (\al_8 + \al_3 + \al_9) g^2 + (\al_1 + \al_2) {g^{\prime}}^2\,,  \\
\delta g^1_Z & =& - \al_3 G^2 \,\hskip 0.2cm {\rm where}\,\, G^2=g^2+{g^\prime}^2\label{tri1}.
\eea
Due to the difference in the definition of the covariant differential operator,
our triple chiral coefficients have extra signs compared 
with those in \cite{Appelquist:1993ka}.
Current precision on TGC allows us to  drop the negligible  
terms induced through the diagonalization and normalization between
Z boson and photon. 

There are no experimental data relaxing  the custodial symmetry 
 except L3 collaboration
\cite{L3group}  from where we take   $\delta  k_Z = -0.076\pm 0.064$ 
as one of the inputs.  Other inputs $\delta  k_\gamma =-0.027 \pm 0.045 $ and 
$\delta  g_Z^1 = -0.016 \pm 0.022$ are taken from LEP Electroweak working 
group \cite{LEPEWWG, Heister:2001qt}. All these data are extracted from one-parameter 
TGC fits as the two-parameter  fits on $\de g^1_Z$ and $\de k_\gamma$  
show larger errors while three parameter fits do not exist.  
We found TGC errors are quite large as reported in D0 collaboration \cite{Abazov:2005ys} at Tevatron.
\par Further the most stringent constraints data from LEP2 
are preferably analyzed  relaxing  the custodial $SU(2)$ gauge 
symmetry as it is natural in the framework
of the EWCL to have a non-vanishing $\al_9$ if the
underlying dynamics break this symmetry explicitly \cite{SSVZ}.  
Each of these data corresponds to a set of solution 
for $\al_2(m_Z)$ , $\al_3(m_Z)$,  $\al_9(m_Z)$ and are 
assumed to be extracted from independent measurements. 
Computing the anomalous TGC in EWCL from these data we get
\begin{equation}
\label{3pfit}
\begin{array}{rl}
\al_2 =  \!\!\!&(-0.09\pm 0.14)\\
\al_3 =  \!\!\!&(+0.03\pm 0.04)\\
\al_9=  \!\!\!&(+0.12\pm 0.12)
\end{array}\!
\rho_{co.}\! \! = \! \! \left(\begin{matrix}
1 \,\,\,& & \cr 
0 & 1 \,\,\,& \cr 
-.7&-.3&1\,\,\,\cr
\end{matrix}\!\!
\right)\!\! .
\end{equation}
Correlations among the experimental observables affects the $\rho_{co.}$ 
insignificantly, without changing their central values.
We observe that $\al_3(m_Z)$ is more tightly constrained 
than $\al_2(m_Z)$ and $\al_9(m_Z)$. Anomalous TGC  are observed to be 
one order more constrained w.r.t. the tree level
unitary bounds from $f_1 {\bar f_2} \rightarrow
V_1 V_2$ at  $\Lambda \geq 1$  (TeV)   \cite{Baur:1987mt}.
\bea
\left\vert\delta k_\gamma\right\vert < \frac{1.86}{ \Lambda^2}\,,
\left\vert\delta k_Z \right\vert <  \frac{0.85}{ \Lambda^2}\,, 
\left\vert\delta g^1_Z \right\vert < \frac{0.87}{ \Lambda^2}\,\,\, .\label{tri-uni}
\eea

The four point chiral coefficients or the quartic 
gauge couplings (QGC) have no experimental data
and usually are assumed to be of order one.
Partial wave unitary bounds of longitudinal vector boson
scattering processes can be used to put bounds on the magnitude
of those chiral coefficients. Absence of Higgs boson or other 
resonances below the UV cutoff $\Lambda$
renders the form factor of these scattering amplitudes to be energy dependent.
We use the following five conditions to constrain five chiral coefficients,
$\al_4$, $\al_5$, $\al_6$, $\al_7$, and $\al_{10}$:
\begin{eqnarray}
&&\hskip -0.5 cm 
\begin{array}{l}
|4 \al_4 + 2\al_5 |  <  {3 \pi} \frac{v^4 }{ \Lambda^4},
\\
| 3 \al_4 + 4 \al_5 |  < {3 \pi } \frac{v^4 }{ \Lambda^4},
\\
|\al_4 + \al_6 + 3 (\al_5 + \al_7) | <  {3 \pi } \frac{v^4 }{ \Lambda^4},
\\
|2 (\al_4 +\al_6) + \al_5  +\al_7 |  <  {3 \pi } \frac{v^4 }{ \Lambda^4},
\\
| \al_4 + \al_5 + 2 (\al_6 + \al_7 + \al_{10}) |  < \,\,\, \frac{6 \pi }{5 } \frac{v^4 }{ \Lambda^4},
\end{array}
\label{quartic-bounds}
\end{eqnarray}
where the bounds are obtained from $W^+_L W^+_L \rightarrow W^+_L W^+_L$,
$W^+_L W^-_L \rightarrow W^+_L W^-_L$, $W^+_L W^-_L \rightarrow Z_L Z_L$,
$W^+_L Z_L   \rightarrow W^+_L Z_L $, and $Z_L Z_L \rightarrow Z_L Z_L$, respectively.

The $\al_1$ and $\al_8$ are small in magnitude and are dropped.
The contributions of TGC and terms proportional 
to $v^2/ \Lambda^2$ are also dropped here,
but are included in the numerical analysis. 
We have avoided a more strict procedure to derive 
unitary bounds as shown in 
\cite{Gounaris:1994cm}.

Above is our current knowledge on those dimensionless chiral
coefficients. Below we analyze how uncertainty in those dimensionless
chiral coefficients can affect the value of 
$S(\Lambda)$-$T(\Lambda)$-$U(\Lambda)$.

In order to determine the values of $S(\Lambda)$-$T(\Lambda)$-$U(\Lambda)$,
we need the RGEs of $\al_1$-$\al_0$-$\al_8$, which are given as
$8 \pi^2 \bigg[d~\alpha_i/ d~t\bigg] = \beta_{\alpha_i}$ while the $\beta_{\alpha_{1,8,0}}$ are 
\bea
\beta_{\al_1} & = &\frac{1}{12}  +  4 \my{\al_1}  g^2 - \my{\al_8}  g^2 \wsep
+ \frac{5  }{2} \my{\al_2} g^2 - 
  \frac{5 }{6}  \my{\al_3} g^2+ 
  \frac{1 }{2} \my{\al_9} g^2 \label{betaa1}
\eea
\bea
\beta_{\al_8}&=&- \frac{\al_0 }{ 2} + 
\my{\al_1} {g'}^2 + 12 \my{\al_8} g^2 \wsep
+  \frac{5  }{6} \my{\al_2} {g'}^2 - 
  \frac{1 }{2} \my{\al_3}  {\my{g'}}^2
+ \frac{17  }{6} \my{\al_9} g^2
\eea
\bwt
\bea
\beta_{\al_0}&=&
-\frac{3 {\my{g'}}^2}{8} + \frac{9 \al_0 g^2 }{ 4} - \frac{9 \al_0 {\my{g'}}^2}{ 4 } + 
  \my{\al_1} \frac{3 g^2 {\my{g'}}^2}{4} - 
  \my{\al_8} \frac{3\,g^4}{8} \wsep  + 
  \my{\al_2} \sbbkf{ \frac{3 g^2 {\my{g'}}^2}{2} - 
     \frac{3 {\my{g'}}^4}{4} } + 
  \my{\al_3} \frac{3  g^2 {\my{g'}}^2}{2} + 
  \my{\al_9} \sbbkf{ -\frac{g^4}{2} + 
     \frac{3 g^2 {\my{g'}}^2}{4} }  \wsep - 
  \my{\al_4} \sbbkf{ \frac{15 g^2 {\my{g'}}^2}{4} + 
     \frac{15 {\my{g'}}^4}{8} }  - 
  \my{\al_5} \sbbkf{ \frac{3 g^2 {\my{g'}}^2}{2} +
     \frac{3 {\my{g'}}^4}{4} } \wsep - 
  \my{\al_6} \sbbkf{ \frac{3 g^4}{4} + 
     \frac{33 G^4}{8} }  - 
  \my{\al_7} \sbbkf{ 3 g^4 + 3 G^4 }    - 
  \my{\al_{10}} \sbbkf{ \frac{9 G^4}{2} }   
\,,
\label{rge1}
\eea
\ewt
where we observe that all TGC contributes to the $\beta_{\al_{0,1,8}}$  
while QGC contributes only  to the $\beta_{\al_0}$. 
This implies QGC do not contribute to the $S$ parameter.
The $S(\Lambda)$, $T(\Lambda)$, and $U(\Lambda)$, are computed from 
 the evolved $\al_1(\Lambda )$-$\al_0(\Lambda)$-$\al_8(\Lambda)$ trough RGE.
The $S(\Lambda)$, $T(\Lambda)$ and $U(\Lambda)$ are  
the values of the parameters $S$, $T$, 
and $U$ at the matching scale $\Lambda$,
where the EWCL matches with 
fundamental theories, Technicolor models, 
extra dimension models, Higgsless models, etc. 

How does the uncertainty of TGC affect the value of $S(\Lambda)$?
To answer this question we set all QGC to zero at $\Lambda=m_Z$ to the study the effect of TGC 
on $S(\Lambda)-T(\Lambda)$  plane which is  depicted in Fig. 1. 
We highlight some features of this figure.

{\bf (1)} In absence of TGC contribution ( red contours ),  $S(\Lambda)$  
becomes more negative as $\Lambda$ increases w.r.t. the reference LEP1 fit 
contour at $\Lambda = m_Z$. This is in agreement with
the observation of Ref. \cite{Bagger} and Ref. \cite{Peskin:2001rw}. 
Inclusion of  TGC contribution as obtained from 
LEP2 fit (Eq. \ref{3pfit}), makes  $S(\Lambda)$ almost unchanged (the solid line).  

\begin{figure}
\begin{center}
\includegraphics[scale=0.35]{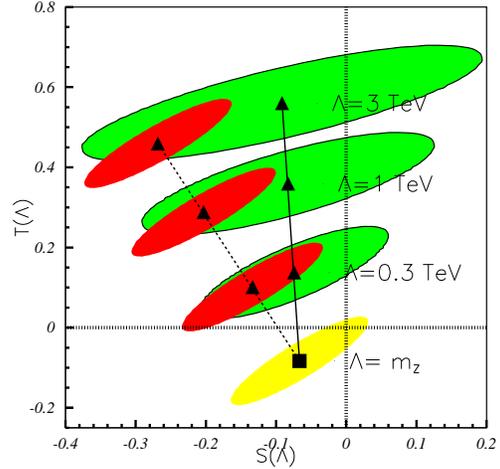}
\end{center}
\caption{ $S(\Lambda)-T(\Lambda)$ contours at $\Lambda=\m_z$, $300$ GeV,
$1$ TeV, and $3$ TeV, respectively. TGC  uncertainty is included in green
contours  while the red contours are without it. } 
\label{fig1}
\end{figure}

{\bf (2)} We observe that when TGC with $1 \sigma$ uncertainty at $\Lambda =m_Z$ 
are taken into account, $S(1$ TeV $)$ can vary between  $-0.3\le S(1\,{\rm  TeV})\le 0.12$
which is almost $3 \sigma$ away from the prediction of $S(1\,\, {\rm TeV})$ 
without these uncertainties. 
Analysis with Tevatron data 
and LEP2 two dimensional TGC fit data 
would exceed this limit dramastically.  
\Black

{\bf (3)} The  TGC contributions can at most lower the value of 
$T(1\, {\rm TeV})$ by  
$\left \vert \Delta T(1\, {\rm  TeV})\right\vert\approx  0.1$. 
Thus the contribution of TGC  is not large enough 
to cancel the large leading contribution 
from $3{g'}^2/8$ in the $\beta$ function of $T$ parameter, which makes $T(\Lambda)$ positive for high energy.
 
Experimental data on the TGC allows the radiative 
mechanism to render large +ve $T(\Lambda)$. To realize 
vanishing $T(\Lambda)$ with QGC switched off would 
require $T(\m_z)$ to be negative $\approx -0.4$ or so, 
which is in confrontation  to the  global fit 
value given in Eq. (\ref{stfit}).  

Whether is it possible to find a solution in the parameter space
of the EWCL? To answer this question, it is worthwhile 
to understand the evolution of  the beta functions 
of QGC affecting  $T(\Lambda )$ parameter. We observe that $\al_4, \al_5$  
terms come along with ${g^{\prime}}^2$, making them one order weaker 
w.r.t. those of $\al_6$, $\al_7$  and $\al_{10}$.
Assuming unitarity bounds on all anomalous QGC  would be of the same 
order and $\al_{10}$ to dominate among the total QGC contribution.
We find that $\left \vert\al_{10}\right \vert$ has 
to be $\ge 0.03$  to switch the sign of $T(1\,\,{\rm TeV})$, 
which is contradictory to the unitary bound given in Eq. (\ref{quartic-bounds})
at $\Lambda=1$ TeV with $v=246$ GeV. 

The reason for the subdominant behavior of QGC couplings with increasing energies can be explained from the Table 1.
We realize that with the increasing
$\Lambda$ the TGC uncertainty  $\de T^{\textrm{TGC}}$ increases
logarithmicly 
while  the QGC uncertainty $\de T^{\textrm{QGC}}$  decreases
rapidly due to the power dependence in the
unitary bounds given in Eq.(\ref{quartic-bounds}).
Consequently it is observed 
that $\de T^{\textrm{QGC}}$ and $\de T^{\textrm{TGC}}$ 
dominates the error of $T(\Lambda)$ below and 
above  $\Lambda < 950$ GeV,
respectively.

\begin{table}[ht]
\begin{center}
\begin{tabular}{|l|  c| c| c| c| }\hline
& & & & \\
 $\Lambda$ & $T(\Lambda)\pm 1\sigma$ &  $\de T_Z$  &  
$\de T^{\textrm{TGC}}$  & $\de T^{\textrm{QGC}}$  \\ 
\hline
& & & &\\
$0.3$ TeV & $0.25\pm 8.91$   &  $\pm 0.14$  & $\pm 0.06$ & $\pm 8.91$ \\
& & & &\\
$0.5$ TeV & $0.29\pm 1.16$   &  $\pm 0.14$  & $\pm 0.08$ & $\pm 1.15$ \\
& & & & \\
$1$ TeV   & $0.40\pm 0.22$   &  $\pm 0.14$  & $\pm 0.12$ & $\pm 0.10$ \\
& & & & \\
$3$ TeV   & $0.60\pm 0.25$   &  $\pm 0.14$  & $\pm 0.17$ & $\pm 0.04$ \\
& & & &\\ \hline
\end{tabular}
\end{center}
\caption{ Values of $T(\Lambda)$ and $1 \sigma$ errors from $\de T_Z$, $\de T^{\textrm{TGC}}$ and 
$\de T^{\textrm{QGC}}$. }
\label{table1}
\end{table}

From Table. 1, we can conclude that in the constrained EWCL 
parameter space with $1 \sigma$ error in TGC and 
with unitary bounds on QGC, it is unlikely to have 
a scenario with vanishing $T(1\,\,$ TeV$)$ 
while keeping $T(m_Z)=-0.08$.
It is worth mentioning that performing the analysis 
with  two-parameter TGC fits 
$\de T^{\textrm{TGC}}$ becomes larger 
while $\de T^{\textrm{QGC}}$ changes insignificantly. 
However, there are  possible ways to evade this situation: {\bf (1)} Lowering the UV 
scale $\Lambda$ down to $700$ GeV or so,
{\bf (2)} Relaxing the error of $T(1$TeV$)$ to $2 \sigma$ or so, and
{\bf (3)} Generating a large enough positive $T(\Lambda)$ from more fundamental
dynamics, as proposed in most Technicolor models when matched with
the effective theory. 

\par We summarize our study and conclude that 
LEP2 data has constrained the anomalous 
TGC (three point chiral coefficients), 
but allows regions where the $S(m_Z$ parameter can be
explained by the radiative corrections of the TGC accompanying
with a positive $S(\Lambda)$.
This letter shows that the 
negative $S(\Lambda)$ parameter problem can be related to 
the loosely constrained large anomalous TGC ($\al_2$ and $\al_9$ ).
With the current experimental and theoretical knowledge, 
TGC's and QGC's uncertainty can 
undermine our prejudice for discarding 
or accepting a specific EWSB model. 
However the upcoming colliders, with higher sensitivity to the TGC, 
can reduce the parameter space and help to
pinpoint the correct model of EWSB.

We would like to thank Ulrich Parzefall for communication on the TGC
measurements at LEP2, and Masaharu Tanabashi for stimulating discussion.
SD  thanks  SERC, DST, India and the core-university program of the JSPS for the 
partial financial support. Work of KH is supported in part 
by MET, Japan. QSY thanks the theory group of physics department, Tsinghua university 
for helpful discussion and thanks JSPS and NCTS (Hsinchu, Taiwan) for partial financial support.

\end{document}